\documentclass[oneside,onecolumn,12pt,a4paper]{article}
\usepackage{graphicx}
\usepackage{dsfont}
\usepackage{natbib}
\usepackage{bbding}
\usepackage{setspace}
\usepackage{braket}
\usepackage{sectsty}
\usepackage{tikz}
\usepackage{bm}
\usepackage{amsmath}
\usepackage{mathtools}
\usepackage{amssymb}
\usepackage{hyperref}
\usepackage{color}
\definecolor{darkblue}{rgb}{0.0,0.0,0.3}
\hypersetup{colorlinks,breaklinks,linkcolor=darkblue,urlcolor=darkblue,anchorcolor=darkblue,citecolor=darkblue}

\def\ci{\perp\!\!\!\perp}

\def\citepos#1{\citeauthor{#1}'s (\citeyear{#1})}

\usepackage{ifthen}
\def\eprinttmp@#1arXiv:#2 [#3]#4@{\ifthenelse{\equal{#3}{}}{\href{http://arxiv.org/abs/#1}{arXiv:#1}}{\href{http://arxiv.org/abs/#2}{arXiv:#2 [#3]}}}
\newcommand{\eprint}[1]{\eprinttmp@#1arXiv: []@}
\newcommand{\doi}[1]{\href{http://dx.doi.org/#1}{doi:#1}}

\sectionfont{\normalfont\normalsize\bfseries}
\subsectionfont{\normalfont\small\bfseries}
\subsubsectionfont{\normalfont\small\bfseries}

\begin{document}

\setstretch{1.2}

\title{Quantum causal models, faithfulness and retrocausality}
  \maketitle

\begin{abstract}
  \citet{WoodSpekkens} argue that any causal model explaining the EPRB correlations and satisfying no-signalling must also violate the assumption that the model faithfully reproduces the statistical dependences and independences---a so-called ``fine-tuning'' of the causal parameters; this includes, in particular, retrocausal explanations of the EPRB correlations. I consider this analysis with a view to enumerating the possible responses an advocate of retrocausal explanations might propose. I focus on the response of \citet{Naeger}, who argues that the central ideas of causal explanations can be saved if one accepts the possibility of a stable fine-tuning of the causal parameters. I argue that, in light of this view, a violation of faithfulness does not necessarily rule out retrocausal explanations of the EPRB correlations, although it certainly constrains such explanations. I conclude by considering some possible consequences of this type of response for retrocausal explanations.
\end{abstract}

\section{Causal models, quantum mechanics and faithfulness}

In the 1960s, \citet{Bell64,Bell66} showed that, given some reasonable assumptions, any local hidden variable model of quantum phenomena must disagree with the predictions of quantum mechanics, a result now known as Bell's theorem. The predictions of quantum mechanics were confirmed in the 1980s \citep{AspectI,AspectII} and, thus, lest Bell's reasonable assumptions be jettisoned, it is widely accepted that certain quantum phenomena cannot be modelled using local hidden variables. But Bell's theorem shows more than just the incompatibility of local hidden variable theories and quantum mechanics; it shows that, regardless of how we model quantum phenomena, quantum mechanics violates a condition that Bell called local causality: the probabilities attached to the values of some quantum variable remain unaffected, given a full specification of the variable's relevant causal past, by events in some spacelike separated region \citep{Norsen09,Norsen11}. Thus it can be argued that Bell's theorem shows us that quantum mechanics is incompatible with this very basic intuitive notion of causality.

Despite this result (or, indeed, because of it), there has been recent interest in the relationship between causal models and quantum mechanics and this has  led to the application of schema designed to permit an inference from observed statistical correlations to causal structure---so-called causal discovery algorithms \citep{Spirtes00,Pearl09}---to the observed statistical correlations of quantum mechanics. According to the causal modelling program, observed statistical dependences and independences between distinct measurable elements of some system can be represented by directed acyclic graphs (DAGs) in which each vertex corresponds to the distinct variables of the system and each edge corresponds to statistical dependence relations between pairs of variables. These patterns of statistical dependences and independences and the corresponding graphical structures can be interpreted as causal Bayesian networks, thereby representing a causal structure between elements of a system. A causal structure delineates a hierarchy of \emph{descendants} (downstream variables), \emph{ancestors} (upstream variables) and \emph{parents} (direct ancestors), while variables with no parents are \emph{exogenous}. A causal model takes a causal structure and specifies a conditional probability distribution for each variable given its parents, $\mathrm{P}(X \mid pa(X))$, which embodies the statistical dependences and independences; the exogenous variables are, by construction, independently distributed.

In order that such causal models have the right sort of features to enable the discovery of causal structure from observed statistical correlations, a series of constraints must be placed on the graphical representations that are general expressions of features that we take to be characteristic of causation. The role of these constraints is to delimit as much as possible the underdetermination in the move from statistical data to causal structure and they serve to clarify the scientific rationale that underlies such causal inference. Two of these constraints are already built into the graphical structures themselves: the causal graphs are directed (expressing causal asymmetry) and acyclic (no causal loops). There are then three central assumptions. The first of these is that the joint distribution over all the variables is constrained to factorise into the product of the conditional probabilities, and from this follows the causal Markov condition: every variable is conditionally independent of its nondescendants given its parents. If we were to use this inference alone to discover the causal structure given some probability distribution, the causal structure would remain significantly underdetermined; an equivalence class of causal structures may in general support a single probability distribution. The second central assumption is that any resultant causal model faithfully reproduces the statistical dependences and independences; that is, every statistical independence implies a causal independence (or, no causal independence is the result of a fine-tuning of the model). Finally, we must assume that the model is causally sufficient; that is, all the common causes of the measured variables have themselves been measured (or, no unmeasured common causes).

When certain quantum phenomena, particularly EPRB correlations, are modelled in this framework it becomes apparent that one or more of the causal assumptions that underlie the framework must be violated. Indeed, the usual response is that the causal Markov condition fails in quantum mechanics \citep{Butterfield92}. This strongly suggests that the quantum correlations cannot be modelled according to such a classical causal framework---which is not entirely surprising given Bell's theorem. Despite this, the quantum causal modelling program is an attempt to capture the quantum correlations within some causal modelling framework and this then requires one or more of the classical aspects of the framework to be appropriately modified. In this context, \citet{WoodSpekkens} have argued that any causal structure that accounts for the quantum correlations must violate the principle of faithfulness, concluding that the no-signalling requirement of the quantum correlations can only be achieved in the causal modelling framework by fine-tuning the parameters in the model.

Since the faithfulness assumption is a central assumption for the discovery of causal structure from observed statistical correlations, the inference from Wood and Spekkens is that the quantum correlations cannot be given a causal explanation \emph{\`{a} la} the causal modelling program. If we were to forego the faithfulness assumption, the intuition would be that not only would we have no way of constraining the equivalence class of causal structures that support a single probability distribution, but we would lose our ability to infer direct causal relations as a result of interventions on the system, since we would have no way of knowing whether such relations were potentially subject to confounding through fine-tuning. Thus, in accordance with the framework of Wood and Spekkens, maintaining the faithfulness assumption leads to rejecting causal explanations in quantum mechanics.

Wood and Spekkens argue that \emph{all} causal explanations of the no-signalling EPRB correlations must violate the faithfulness assumption, and therefore should be ruled out as viable explanations. This current work is interested in particular in the fate of retrocausal explanations, and proceeds as follows.

I begin in Section~\ref{sec:nofinetune} by outlining Wood and Spekkens' argument that there is an insurmountable tension between the no fine-tuning assumption, the no signalling requirement and a causal explanation of the EPRB correlations. I then introduce in Section~\ref{sec:primretro} an explicit example of a basic retrocausal model to illustrate to what exactly fine-tuning amounts. Section~\ref{sec:posresp} briefly considers some arguments against the no fine-tuning assumption: that fine-tuning is ubiquitous and not a problem \citep{Cartwright01}; that fine-tuning that is independently justified is not a problem \citep{EggEsfeld}; and that fine-tuning that is stable to disturbances is not a problem \citep{Naeger}. It is this latter response on which I focus, with a view to sketching a retrocausal picture in Sections~\ref{sec:robretro} and~\ref{sec:cancelpath} that seems to satisfy the stability conditions required of the view. The main argument of this paper is that thinking of the fine-tuning of the probabilities associated with EPRB correlations as being grounded in cancelling probability amplitudes---as per the Feynman path integral approach to quantum mechanics---allows the requisite stability to circumvent the conclusions of Wood and Spekkens.

As a concluding discussion, I then speculate in Section~\ref{sec:retromodel} how (or indeed if) this retrocausal picture could map back onto the causal modelling project. In terms of the analysis of Wood and Spekkens, there are two general ways to do this: by accepting that stable fine-tuning is an appropriate addition to the causal modelling project such that a causal model of the EPRB experiment is possible, contra the conclusion of Wood and Spekkens; or by accepting that the retrocausal picture requires sufficiently radical changes to the causal modelling project such that the associated explanation of the EPRB correlations is no longer a causal explanation, thus aligning in spirit with Wood and Spekkens' conclusion that no causal explanation is possible. Either way, the Wood and Spekkens analysis does not sound the death knell for all retrocausal approaches to quantum mechanics.\footnote{To be fair, Wood and Spekkens do qualify that their analysis rules out retrocausal models that can be represented by \emph{directed acyclic} graphs, and it is debatable whether the retrocausal picture outlined in Section~\ref{sec:robretro} can be represented as such. More on this in Section~\ref{sec:quantretro}.}

\section{The ``no fine-tuning'' requirement}
\label{sec:nofinetune}

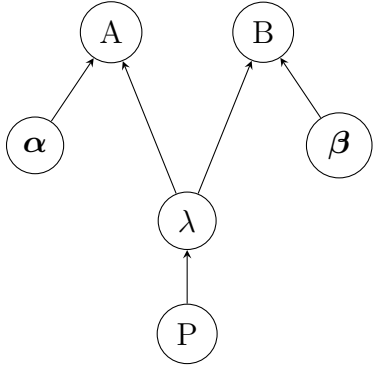
\begin{figure}
  \centering
  \begin{tikzpicture}[scale=1]
    \node (X) at (0,1) [circle,draw=black] {$\bm{\alpha}$};
    \node (A) at (1,2.5) [circle,draw=black] {A};
    \node (B) at (3,2.5) [circle,draw=black] {B};
    \node (Y) at (4,1) [circle,draw=black] {$\bm{\beta}$};
    \draw [-stealth] (X) -- (A);
    \draw [-stealth] (Y) -- (B);
    \node (L) at (2,0) [circle,draw=black] {$\lambda$};
    \node (P) at (2,-1.5) [circle,draw=black] {P};
    \draw [-stealth] (P) -- (L);
    \draw [-stealth] (L) -- (A);
    \draw [-stealth] (L) -- (B);
  \end{tikzpicture}
  \caption{Common cause structure of the EPRB experiment.}
  \label{fig:ccbell}
\end{figure}

Consider, as is usual, an EPRB experiment consisting of an entangled bipartite quantum system $\Ket{\psi}$ in the singlet state and where each of the parts of the bipartite state are subject to a local measurement---the settings variable denoted by $\bm{\alpha}$ on one side and $\bm{\beta}$ on the other---with two possible results for each measurement---the outcome variable denoted by $A$ on the $\bm{\alpha}$ side and $B$ on the $\bm{\beta}$ side. According to Bell's theorem, $\mathrm{P}(A,B \mid \bm{\alpha},\bm{\beta})$ is constrained by the impossibility of a causal influence between the two sides of the experiment---Bell's notion of local causality \citep{Norsen11}---leading to the assumption that any correlation must be due to a common cause $\lambda$. This can be depicted as a DAG, as in Fig.~\ref{fig:ccbell} (where the variable $P$ consists of state preparations like $\Ket{\psi}$).

Applying the causal Markov condition to this graph gives us the following conditional independences:\footnote{$(X \ci Y \mid Z)$ denotes that $X$ and $Y$ are conditionally independent given $Z$:
\begin{equation*}
  \mathrm{P}(X,Y \mid Z) = \mathrm{P}(X \mid Z)\cdot\mathrm{P}(Y \mid Z).
\end{equation*}}
\begin{align*}
  (A \ci \bm{\beta},B \mid \bm{\alpha},\lambda)\\
  (B \ci \bm{\alpha},A \mid \bm{\beta},\lambda)
\end{align*}
which is just a statement of Bell's local causality. From this condition, and the assumption that there are no correlations between the settings and the hidden variables, $(\bm{\alpha} \ci \bm{\beta},\lambda)$ and $(\bm{\beta} \ci \bm{\alpha},\lambda)$, it follows that $\mathrm{P}(A,B \mid \bm{\alpha},\bm{\beta})$ must satisfy the Bell inequalities. That there be no superluminal signalling also follows from Fig.~\ref{fig:ccbell}, $(\bm{\alpha} \ci B \mid A)$ and $(\bm{\beta} \ci A \mid B)$.

Wood and Spekkens consider this problem in reverse, as per the casual discovery schema. Rather than attempting to explain the EPRB correlations using causal structure, as Bell attempted, the casual discovery schema can be employed to illuminate the possible causal structures from the observed statistical correlations of quantum mechanics. The observed statistical correlations are, of course, marginal independence of the settings, $(\bm{\alpha} \ci \bm{\beta})$, and no-signalling, $(A \ci \bm{\beta} \mid \bm{\alpha})$ and $(B \ci \bm{\alpha} \mid \bm{\beta})$. When the no-signalling constraint is combined with the faithfulness assumption---that every statistical independence implies a causal independence---we infer that there can be no (direct or mediated) causal influences from one side of the EPRB experiment to the other (which is just Bell's notion of local causality). Again, it follows from this that $\mathrm{P}(A,B \mid \bm{\alpha},\bm{\beta})$ must satisfy the Bell inequalities, and thus the fact that the Bell inequalities are violated by the EPRB correlations renders such causal discovery defective.

Apart from concluding that this set of conditional independences is ``too impoverished'' for some prominent causal discovery algorithms to lead to significant insight into the casual structure of quantum mechanics, Wood and Spekkens demonstrate a fundamental tension between the no-signalling constraint, the faithfulness assumption and the possibility of a causal explanation.  More precisely, they show that the following three assumptions form an inconsistent set:
\begin{enumerate}
  \item The predictions of quantum theory are correct---that is, the conditional independences $(\bm{\alpha} \ci \bm{\beta})$, $(A \ci \bm{\beta} \mid \bm{\alpha})$ and $(B \ci \bm{\alpha} \mid \bm{\beta})$ are satisfied, and a Bell inequality is violated;
  \item The observed statistical dependences and independences can be given a causal explanation as per the causal modelling program;
  \item The faithfulness assumption holds---that is, there is no fine-tuning.
\end{enumerate}
Wood and Spekkens advocate dropping the second assumption, the contrapositive of which is that any purported causal explanation of the EPRB correlations falls afoul of the tension between the no-signalling constraint (first assumption) and no fine-tuning (third assumption), including retrocausal explanations. Let us consider a basic retrocausal model to explore this tension.

\subsection{Fine-tuning in a retrocausal model}
\label{sec:primretro}

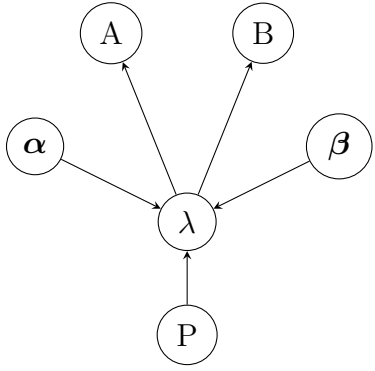
\begin{figure}
  \centering
  \begin{tikzpicture}[scale=1]
    \node (X) at (0,1) [circle,draw=black] {$\bm{\alpha}$};
    \node (A) at (1,2.5) [circle,draw=black] {A};
    \node (B) at (3,2.5) [circle,draw=black] {B};
    \node (Y) at (4,1) [circle,draw=black] {$\bm{\beta}$};
    \node (L) at (2,0) [circle,draw=black] {$\lambda$};
    \node (P) at (2,-1.5) [circle,draw=black] {P};
    \draw [-stealth] (P) -- (L);
    \draw [-stealth] (L) -- (A);
    \draw [-stealth] (L) -- (B);
    \draw [-stealth] (X) -- (L);
    \draw [-stealth] (Y) -- (L);
  \end{tikzpicture}
  \caption{Causal relations in a basic retrocausal model.}
  \label{fig:primretro}
\end{figure}

Consider the retrocausal model depicted in Fig.~\ref{fig:primretro}. In this model, the measurement settings have a direct causal influence on the hidden variable $\lambda$ and only an indirect cause on the outcomes $A$ and $B$. We can imagine the model to work in the following way. We begin with an entangled bipartite quantum system $\Ket{\psi}$, a pair of photons in a singlet state: %spin-$\frac{1}{2}$ particles in a spin singlet state:
\begin{equation*}
  \Ket{\psi} = \frac{1}{\sqrt{2}}(\Ket{+-} - \Ket{-+})
\end{equation*}
In a typical CHSH test, each of the pair of photons is subject to one of two different local measurement---on the $\bm{\alpha}$ side we randomly choose to measure $\alpha_{1}$ or $\alpha_{2}$, and on the $\bm{\beta}$ side $\beta_{1}$ or $\beta_{2}$---with two possible results for each measurement, i.e. $A = \{+,-\}$, $B = \{+,-\}$. We observe that the outcomes are correlated with probability $\sin^{2}{\frac{\theta_{ij}}{2}}$ and anti-correlated with probability $\cos^{2}{\frac{\theta_{ij}}{2}}$ (where $\theta_{ij}$ is the angle between the measurement settings, $i = \alpha_{1},\alpha_{2}$ and $j = \beta_{1},\beta_{2}$).

%in one direction out of three coplanar axes, any two of them separated by an angle $\theta_{ij} = \frac{2\pi}{3}$ ($\bm{\alpha} = \{\alpha_{1},\alpha_{2},\alpha_{3}\}$, $\bm{\beta} = \{\beta_{1},\beta_{2},\beta_{3}\}$). The result of any measurement is either spin up, $+$, or spin down, $-$ ($A = \{+,-\}$, $B = \{+,-\}$). We observe the following correlations: when the measurement settings are along the same direction (i.e. $\alpha_{i}$ and $\beta_{j}$ with $i = j$) then we have perfect anti-correlation of the outcomes (i.e. $A = +$ and $B = -$ or $A = -$ and $B = +$); averaged over all the measurement settings, we see that the outcomes are random (i.e. $A = B$ half the time, and $A \neq B$ half the time).

We can provide a retrocausal model of this experiment in a straightforward way: let $\lambda$ be constituted by a pair of ``beables'' that each encode information about their respective measurement outcome as a function of the setting (and together encode information about the joint measurement outcome as a function of the settings). Thus, for settings, say, $\alpha_{1}$ at $\bm{\alpha}$ and $\beta_{2}$ at $\bm{\beta}$, we could represent the possible values of $\lambda$ as $(+_{\alpha_{1}},+_{\beta_{2}})$, $(+_{\alpha_{1}},-_{\beta_{2}})$, $(-_{\alpha_{1}},+_{\beta_{2}})$ or $(-_{\alpha_{1}},-_{\beta_{2}})$. To achieve the observed quantum probabilities, these values of $\lambda$ need to occur in accordance with the observed probabilities; that is:
\begin{alignat*}{2}
  \mathrm{P}(+_{i},+_{j} \mid \bm{\alpha},\bm{\beta}) &= \mathrm{P}(-_{i},-_{j} \mid \bm{\alpha},\bm{\beta}) = \frac{1}{2}\sin^{2}{\frac{\theta_{ij}}{2}}\\
  \mathrm{P}(+_{i},-_{j} \mid \bm{\alpha},\bm{\beta}) &= \mathrm{P}(-_{i},+_{j} \mid \bm{\alpha},\bm{\beta}) = \frac{1}{2}\cos^{2}{\frac{\theta_{ij}}{2}}
\end{alignat*}

% = \frac{3}{8}
% = \frac{1}{8}
%  \mathrm{P}(+_{i},+_{i} \mid \bm{\alpha},\bm{\beta}) = & \mathrm{P}(-_{i},-_{i} \mid \bm{\alpha},\bm{\beta}) = 0\\
%  \mathrm{P}(+_{i},-_{i} \mid \bm{\alpha},\bm{\beta}) = & \mathrm{P}(-_{i},+_{i} \mid \bm{\alpha},\bm{\beta}) = \frac{1}{2}\\

Let us check to see whether we achieve the no-signalling constraint on this casual model. Consider $(A \ci \bm{\beta} \mid \bm{\alpha})$: the conditional probability distribution over the outcomes at $A$ is independent of the setting $\bm{\beta}$ given the setting $\bm{\alpha}$. Let us take the value of $\bm{\alpha} = \alpha_{2}$ and check whether changing the setting at $\bm{\beta}$ will alter the conditional probability distribution of outcomes at $A$. First, take $\bm{\beta} = \beta_{1}$: $\lambda$ could be any one of $(+_{\alpha_{2}},+_{\beta_{1}})$, $(+_{\alpha_{2}},-_{\beta_{1}})$, $(-_{\alpha_{2}},+_{\beta_{1}})$ or $(-_{\alpha_{2}},-_{\beta_{1}})$, which arise with probability $\frac{1}{2}\sin^{2}{\frac{\theta_{ij}}{2}}$, $\frac{1}{2}\cos^{2}{\frac{\theta_{ij}}{2}}$, $\frac{1}{2}\cos^{2}{\frac{\theta_{ij}}{2}}$ and $\frac{1}{2}\sin^{2}{\frac{\theta_{ij}}{2}}$, respectively. Thus, since
\begin{alignat*}{2}
  \mathrm{P}(A = +) &= \mathrm{P}(+_{\alpha_{2}},+_{\beta_{1}}) + \mathrm{P}(+_{\alpha_{2}},-_{\beta_{1}})
  \shortintertext{and}
  \mathrm{P}(A = -) &= \mathrm{P}(-_{\alpha_{2}},+_{\beta_{1}}) + \mathrm{P}(-_{\alpha_{2}},-_{\beta_{1}}),
  \shortintertext{then}
  \mathrm{P}(A = +) &= \mathrm{P}(A = -) = \frac{1}{2}\sin^{2}{\frac{\theta_{ij}}{2}} + \frac{1}{2}\cos^{2}{\frac{\theta_{ij}}{2}} = \frac{1}{2}.
\end{alignat*}

Now take $\bm{\beta} = \beta_{2}$: $\lambda$ could be any one of $(+_{\alpha_{2}},+_{\beta_{2}})$, $(+_{\alpha_{2}},-_{\beta_{2}})$, $(-_{\alpha_{2}},+_{\beta_{2}})$ or $(-_{\alpha_{2}},-_{\beta_{2}})$, which arise with probability $\frac{1}{2}\sin^{2}{\frac{\theta_{ij}}{2}}$, $\frac{1}{2}\cos^{2}{\frac{\theta_{ij}}{2}}$, $\frac{1}{2}\cos^{2}{\frac{\theta_{ij}}{2}}$ and $\frac{1}{2}\sin^{2}{\frac{\theta_{ij}}{2}}$, respectively. Thus
\begin{equation*}
  \mathrm{P}(A = +) = \mathrm{P}(A = -) = \frac{1}{2}\sin^{2}{\frac{\theta_{ij}}{2}} + \frac{1}{2}\cos^{2}{\frac{\theta_{ij}}{2}} = \frac{1}{2}.
\end{equation*}
We see then that the conditional probability distribution over the outcomes at $A$ is the same regardless of which setting at $\bm{\beta}$ we choose. We could similarly choose $\bm{\alpha} = \alpha_{1}$ and achieve the same result. This is just the conditional independence of $A$ and $\bm{\beta}$ given $\bm{\alpha}$.

% And finally take $\bm{\beta} = \beta_{3}$: $\lambda$ could be any one of $(+_{3},+_{3})$, $(+_{3},-_{3})$, $(-_{3},+_{3})$ or $(-_{3},-_{3})$, which arise with probability 0, $\frac{1}{2}$, $\frac{1}{2}$ and 0, respectively. Thus $\mathrm{P}(A = +) = \mathrm{P}(A = -) = 0 + \frac{1}{2} = \frac{1}{2}$.

However, we have had to ``tune'' the conditional probability distributions (the so-called ``causal-statistical parameters'' of \citet[p.~4]{WoodSpekkens}) in order to avoid the possibility of signalling. If these probabilities were to change just slightly, this would destroy the finely balanced independence of the measurement setting at $\bm{\beta}$ and the outcome at $A$.\footnote{Of course, in a maximally entangled state, as with $\Ket{\psi}$ given above, the symmetry of the system renders certain other independences that belie the causal model given in Fig.~\ref{fig:primretro}, including $(A \ci \bm{\alpha})$ and $(B \ci \bm{\beta})$ \citep[p.~8]{Naeger}. If $\Ket{\psi}$ were to slightly deviate from maximal entanglement these independences would become dependences as per Fig.~\ref{fig:primretro} and yet the system could still violate the Bell inequalities.} That is, some small change in the conditional probability distributions would reveal, in terms of this example, which measurement setting were chosen at $\bm{\beta}$. (For instance, $\bm{\beta} = \beta_{1}$ might give us $\mathrm{P}(A = +) = \frac{3}{5}$ while $\bm{\beta} = \beta_{2}$ might give us \linebreak $\mathrm{P}(A = +) = \frac{2}{5}$.) Thus we can see explicitly the problem faced by causal explanations of the EPRB correlations. While $\mathrm{P}(A,B \mid \bm{\alpha},\bm{\beta})$ suggests that there is a correlation that needs to be represented in a causal model of the EPRB experiment, the no-signalling constraint and the faithfulness assumption prohibit any appropriate connection between the relevant variables of the model.

\section{Possible responses}
\label{sec:posresp}

Wood and Spekkens conclude that the EPRB correlations simply resist any causal explanation and advocate a replacement of the formalism of classical probability theory with some noncommutative generalisation (as per \citet{LeiferSpekkens2013}). While this is certainly a clear direction out of the logical space circumscribed by their analysis, it is not the only possible solution. Another open avenue, and one more amenable to causal explanations of the EPRB correlations, is to question the strength of the faithfulness assumption. After all, Wood and Spekkens justify maintaining this assumption on the ground that it is a central assumption of causal discovery algorithms. (In defence of this justification, we can imagine the motivation behind causal discovery algorithms as an attempt to represent the process by which an agent might make causal inferences about the world. And it would seem there is a sort of ``Occam's razor'' principle when an agent makes such inferences---agents do not usually expect there to be causal relations where they see no statistical correlations.\footnote{\citepos{Pearl09} illustration of this is that we do not expect a picture of a chair to be a picture of two chairs positioned such that one hides the other \citep[p.~9]{WoodSpekkens}.})

But as with all such principles, it would be a surprise if such an expectation of causal inference were exceptionless. Indeed, \citet[p.~244]{Cartwright01} makes the claim that there are many cases in sociology, economics and medicine in which significant caveats are attached to the use of the faithfulness assumption. In such cases, we usually have evidence of dependences and independences independent of the statistical observations which we can use to guide any modelling of the causal relations in the system. When we lack independent evidence, \citet[p.~254]{Cartwright01} asks, what reason do we have for assuming that faithfulness holds?
\begin{quote}
  \singlespace
  \upshape
  \small
  \sffamily
  Where we don't know, we don't know. When we have to proceed with little information we should make the best evaluation we can for the case at hand---and hedge our bets heavily; we should not proceed with false confidence having plumped either for or against some specific hypothesis---like faithfulness---for how the given system works when we really have no idea.
\end{quote}

Another option available to the advocate of a causal explanation of the EPRB correlations is to provide independent motivation for why there is fine-tuning in some particular case. Such an option is taken up by \citet[p.~190]{EggEsfeld} when they attempt to defend Bohmian mechanics---which provides a causal model of the EPRB correlations---from the challenge presented by Wood and Spekkens. They claim that Bohmian mechanics does indeed avoid the possibility of superluminal signalling due to fine-tuning, and this fine-tuning is a consequence of the \emph{quantum equilibrium hypothesis}, which proposes that the configurations of Bohmian mechanics are randomly distributed according to the distribution given in the Born rule. Insofar as the quantum equilibrium hypothesis is justifiable on independent grounds (which itself is controversial, although see \citet{Dürr12} for a defence) we have a justification for why we expect there to be fine-tuning in specific quantum scenarios.

%Dowe suggests some conditions that need to be met for the assumption of faithfulness to be justified, and then argues that in the case of the EPRB experiment, these conditions are not met. The conditions are:
%\begin{enumerate}
%  \item we have reason to suspect causation is symmetric;
%  \item we have a mechanism;
%  \item we think there's no confounding happening.
%\end{enumerate}

There is, however, one possible response that I would like to focus on here, due to \citet{Naeger}. The essence of this response is that the unfaithfulness displayed in the casual relations of some system occurs in a \emph{stable} way. Once we have accounted for some unfaithful independence by identifying the existence of a fine-tuning of the conditional probability distributions of some system, any change in background conditions that disturbs these distributions would transform this unfaithful independence into a faithful dependence (as we saw at the end of Section~\ref{sec:primretro}). Since in the case of EPRB correlations we see the no-signalling constraint upheld in a range of conditions (indeed, its violation has never been measured), we would require in this case a mechanism by which the unfaithful independence is stable under an equivalent range of conditions. This is what N\"{a}ger proposes.

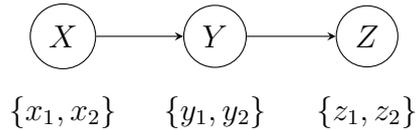
\begin{figure}
  \centering
  \begin{tikzpicture}
    \node (X) at (0,0) [circle,draw=black] {$X$};
    \node at (0,-1) {$\{x_{1},x_{2}\}$};
    \node (Y) at (2,0) [circle,draw=black] {$Y$};
    \node at (2,-1) {$\{y_{1},y_{2}\}$};
    \node (Z) at (4,0) [circle,draw=black] {$Z$};
    \node at (4,-1) {$\{z_{1},z_{2}\}$};
    \draw [-stealth] (X) -- (Y);
    \draw [-stealth] (Y) -- (Z);
  \end{tikzpicture}
  \caption{Causal sequence with multiple specific values.}
  \label{fig:canpaths}
\end{figure}

\citet[p.~22]{Naeger} suggests that the fine-tuning mechanism in quantum mechanics is what he calls \emph{internal cancelling paths}. This mechanism is analogous to the usual cancelling paths scenario (of the sort reprised by \citet[p.~246]{Cartwright01}) in which there are two or more causal routes between a pair of variables which act to exactly counterbalance each other. In contrast to the usual case, however, the mechanism of internal cancelling paths does not manifest at the level of variables, but at the level of values. If we imagine that multiple specific values of a causal sequence of variables, $X$, $Y$, $Z$, as in Fig.~\ref{fig:canpaths}, are pairwise related by conditional probabilities, $\mathrm{P}(y_{i} \mid x_{i})$, $\mathrm{P}(z_{i} \mid y_{i})$, and further imagine that when these probabilities combine along the sequence from the initial to the final variable all such sequentially combined probabilities lose their dependence upon the initial pairwise conditional probabilities ($\mathrm{P}(z_{i} \mid x_{i})$ is independent of $i$), then we can see how it is possible for the final variable to lose its dependence on the value of the initial variable. Thus, rather than a balance being achieved between two or more causal routes connecting a pair of variables, in the case of internal cancelling paths we have a balance being achieved between two or more sequences of pairwise conditional probabilities connecting specific values of the variables. When the probabilities are right, the dependence of the final variable on the initial variable vanishes.

For \citet[p.~22]{Naeger}, the EPRB correlations display just such a mechanism, where the first variable of the sequence is the entangled state prior to collapse, the second is a collapsed state and the third is the corresponding measurement outcome. Given suitable specifications for the conditional probabilities between the values of these variables, the probability of a measurement outcome is rendered independent of the value of the entangled state prior to collapse. But the specific details and mechanism of N\"{a}ger's treatment need not concern us here. What is important for him in this case is that these pairwise conditional probabilities are stable to disturbances of the system. Thus ``if the laws of nature are such that disturbances always alter the different paths in a balanced way, then it is physically impossible to unbalance the paths'' \citep[p.~26]{Naeger}.

%(for the maximally entangled state in the example in Section~\ref{sec:primretro}, the probability of the measurement outcome is $\cos^{2}{\frac{\theta_{ij}}{2}}$, which is independent of the value of \lambda$$)

This then shapes as a nuanced response to the claim that the no-signalling constraints and the principle of faithfulness rule out a causal explanation \emph{\`{a} la} Wood and Spekkens. One of the most significant problems with violations of faithfulness is that such violations ultimately undermine our ability to make suitable inferences of causal independence based on statistical independence. It seems though that N\"{a}ger's position circumvents such problems by allowing only a specific kind of unfaithfulness---a principled or law based unfaithfulness that is stable to background conditions. Whatever may be the mechanism in the EPRB experiment, any violation of faithfulness must be universal (since we have never measured any signalling between the two sides of the experiment). The sorts of violations of faithfulness that are addressed by \citet[p.~246]{Cartwright01} can be identified as, say, ordinary path cancellation when we have some independent evidence of a causal mechanism at work; such evidence may allow us to expose the elemental causal relations that cancel in normal circumstances. Since there appears to be some in-principle lack of epistemic access to the quantum state between preparation and measurement, we then lack epistemic access to any purported causal mechanism underlying the EPRB correlations. Thus we could never use such a mechanism to expose any violation of faithfulness (by, say, revealing the elemental values of the internal paths that cancel), and this is of course just a statement that the violation of faithfulness is stable.

The possibility of a stable violation of faithfulness portends the possibility of providing a more detailed retrocausal model that employs such a mechanism. Let us consider what such a model would look like now.

\section{Quantum causal models and retrocausality}
\label{sec:quantretro}

Interestingly, N\"{a}ger chooses the values of his internal cancelling paths to be conditional probabilities. A more promising candidate for such values would be not probabilities but probability amplitudes. This is more promising because we already know of an interpretation of quantum mechanics that employs cancelling sums over probability amplitudes---Feynman's path integral formulation. What is more, there already exists a retrocausal picture of quantum mechanics based on this formulation.

It might be enough at this stage to simply point to this possibility as a demonstration that there is another viable avenue leading us out of the logical space circumscribed by the analysis of Wood and Spekkens. However, I would like to attempt to fill in some of the detail of this alternative path---in Sections~\ref{sec:robretro} and~\ref{sec:cancelpath} I set out a more detailed retrocausal picture and outline a possible blueprint of how cancelling probability amplitudes might work for the EPRB experiment---and speculate in Section~\ref{sec:retromodel} as to how this might map back onto the causal modelling program. I do not have a new causal modelling framework to present here, but I contend that the current causal modelling framework is too blunt an instrument to accommodate this retrocausal picture.

\subsection{A more detailed retrocausal model}
\label{sec:robretro}

\citet{Wharton10b} (with further development in \citet*{WhartonMillerPrice} and \citet*{EvansPriceWharton}) argues for a retrocausal picture of quantum mechanics based on Hamilton's principle and the symmetric constraint of both initial and final boundary conditions to construct equations of motion from a Lagrangian. Wharton treats external measurements as physical constraints imposed on a system in the same way that boundary constraints are imposed on the action integral of Hamilton's principle. Focussing solely on classical fields, Wharton argues that constraining such fields at both an initial and a final temporal boundary (or a closed hypersurface in spacetime) generates two strikingly quantum features: quantisation of the field and contextuality of the unknown parameters characterising the field between the boundaries.\footnote{That there are unknown parameters before the imposition of the final condition is ensured due to the underdetermination of the classical field by the initial data.} Such contextuality of the ``hidden'' parameters is precisely the feature we find inherent in the EPRB experiment.

Thus we get the following picture: a classical field constrained at both an initial and a final temporal boundary permits by construction variables that are correlated with the future measurement of the system; the final measurement does not simply reveal preexisting values of the parameters, but \emph{constrains} those values (just as the initial boundary condition would). From within Wharton's picture, an invariant joint probability distribution associated with each possible pair of initial and final conditions can be constructed \citep[p.~318]{Wharton10}, and the usual conditional probabilities can be formed by conditioning on any chosen portion of the boundary \citep[p.~280]{Wharton10b}. Probability is then a manifestation of our ignorance: if we only knew the initial boundary, we would only be able to describe the subsequent field probabilistically (due to the future constraint); given the final boundary as well, we would then be able to retrodict the field values between the two boundaries.

As \citet[p.~281]{Wharton10b} points out, this picture maps nicely to the Feynman path integral representation of joint probabilities. According to this scheme, the joint probability of particular initial and final state pairs is given by an integral over the classical action \citep[p.~526]{WhartonMillerPrice}:
\begin{equation}
  \label{eq:classfeyn}
  P(\psi_i,\psi_f)=\left|\int \int \left( \sum_{x,t_{i}\rightarrow y,t_{f}} e^{iS/\hbar} \right) \psi^{\ast}_{f}(y) \psi_{i}(x) dx dy \right|^2,
\end{equation}
where $S = \int L dt$.\footnote{And this scheme extends to QFT as a functional integral, again of the classical action:
\begin{equation*}
  \label{eq:quantfeyn}
  P(\phi_{i},\phi_{f})=\left|\int {\cal D}\phi e^{iS[\phi]/\hbar}\right|^2.
\end{equation*}} Here we have a (spatiotemporally defined) action constrained between two temporal boundaries determining the behaviour of a system between the boundaries.

Of course this is only a general retrocausal framework and not a theory of quantum mechanics. We can, however, get an idea of how to apply (\ref{eq:classfeyn}) within the retrocausal framework by considering the example of the Mach-Zehnder interferometer from \citet[p.~529]{WhartonMillerPrice}. Since such a system has highly localised initial and final conditions, and sufficient redundancy in the set of possible paths through the experiment, the expression for the joint probability of particular input/output pairs simplifies to a sum of amplitudes representing coarse-grained classical paths through the system with different phases. Where amplitudes representing particular paths through the interferometer cancel, the inference is made that those paths contain nothing ``ontological''. (Of course, Feynman warned against this sort of ontological interpretation of the path integral formulation because the path integral sums complex amplitudes and not classical probabilities. However, the main argument of \citet{WhartonMillerPrice} is to employ an action symmetry as a justification for just such an ontological interpretation.) As far as this interpretation goes, such empty paths---and indeed the non-empty paths---are constrained to be as such by \emph{both} the initial and final conditions constraining the system, as per (\ref{eq:classfeyn}) (after all, joint probabilities do not distinguish past from future).

While \citet{WhartonMillerPrice} go on to consider a Bell inequality violating system based on the Mach-Zehnder interferometer, a treatment of an EPRB experiment employing the action symmetry of Wharton's retrocausal picture can be found in \citet{EvansPriceWharton}. As a consequence of this action symmetry, it is argued that we should expect the quantum state of the system between preparation and measurement in an EPRB experiment to be at least partly characterised by beables encoding information about the future measurement settings. This then circumvents Bell's theorem by including an explicit dependence of any hidden variables on the future measurement settings and, furthermore, provides an ``action-by-contact'' explanation of the EPRB correlations (as we expect of retrocausal models). \citeauthor{EvansPriceWharton}, however, do not provide an explicit account of cancelling probability amplitudes for the EPRB experiment (as \citet[p.~530]{WhartonMillerPrice} do for the Bell inequality violating Mach-Zehnder interferometer) and thus they overlook an account of how the no-signalling constraint might be met within Wharton's retrocausal picture.

\subsection{A suggestion for cancelling paths}
\label{sec:cancelpath}

One suggestion for how such an explanation of the no-signalling constraint might look, however, can be found in a model of an EPRB experiment based on the same principle of cancelling probability amplitudes.\footnote{With thanks to Gerard Milburn for the idea of this model and the formal detail. There are similarities between this approach and that of \citet{Tucci95}.} According to this model, we simulate making an intermediary joint measurement on the entangled pair of an EPRB experiment that facilitates modelling each side of the experiment along the lines of Fig.~\ref{fig:canpaths}, where $Y$ is the intermediary measurement. Upon doing so, we can reinterpret the joint probability amplitude, $\mathcal{A}_{\alpha\beta}(a,b \mid \psi)$, for a particular pair of outcomes, $a$ and $b$, as a result of some measurements $\alpha$ and $\beta$ in an EPRB experiment as being comprised of a sum of pairs of such amplitudes: one of which, $\mathcal{A}_{\alpha^{\prime}\beta^{\prime}}(\mu,\nu \mid \psi)$, representing a transition from the initial state of the system, $\psi$, into an eigenstate of an arbitrary intermediary measurement, $\alpha^{\prime}$ and $\beta^{\prime}$; and the other of which, $\mathcal{A}_{\alpha\beta}(a,b \mid \mu,\nu)_{\alpha^{\prime}\beta^{\prime}}$, representing a transition from this intermediary state into the final state of the system, where we then sum over the outcomes of the intermediary measurement, $\mu$ and $\nu$, to erase any information gained about the state (so long as the operators corresponding to the intermediary measurement settings form a complete set of projectors).\footnote{This is, of course, just an abbreviated application of the composition law that \citet[p.~115]{Dirac33} and then \citet[p.~27]{Feynman42} employ to derive the path integral approach to quantum mechanics.}

Employing the EPRB setup from Sec.~\ref{sec:primretro}, with measurement settings at $\bm{\alpha}$ and $\bm{\beta}$ chosen to be $\alpha_{1}$ and $\beta_{1}$ respectively, we are free to make a novel choice for these arbitrary intermediary measurements as the unmeasured settings from each side of the experiment, i.e. $\alpha_{2}$ and $\beta_{2}$. Since we are summing over the outcomes of the intermediary measurement---of which there are four possibilities---the joint probability amplitude for the actual outcomes is a sum of four amplitude pairs, the first of each pair (corresponding to the joint outcome of the intermediary measurements) (i) factorises into two amplitudes (a sort of locality condition) and (ii) can be straightforwardly evaluated given the geometry of the experiment:\footnote{For instance, given a standard geometry for obtaining the CHSH inequalities, the values of the factors $\mathcal{A}_{\alpha_{1} \mid \alpha_{2}}(a \mid \mu)$ and $\mathcal{A}_{\beta_{1} \mid \beta_{2}}(b \mid \nu)$ are exclusively from the set $\{\pm\frac{1}{\sqrt{2}}\}$.}
\begin{equation}
  \label{eq:probamp}
  \mathcal{A}_{\alpha_{1}\beta_{1}}(a,b \mid \psi) = \sum_{\mu,\nu=\pm 1}\mathcal{A}_{\alpha_{1} \mid \alpha_{2}}(a \mid \mu) \mathcal{A}_{\beta_{1} \mid \beta_{2}}(b \mid \nu) \mathcal{A}_{\alpha_{2}\beta_{2}}(\mu,\nu \mid \psi),
\end{equation}
where $\psi$ is the initial state, $a$ and $b$ are the outcomes on the $\bm{\alpha}$ and $\bm{\beta}$ sides of the experiment, respectively, and $\mu$ and $\nu$ are the respective intermediary outcomes.

The result of all this is that the joint probability amplitude for the actual outcomes, $\mathcal{A}_{\alpha_{1}\beta_{1}}(a,b \mid \psi)$, is a sum of the amplitudes corresponding to the unmeasured measurement settings, $\mathcal{A}_{\alpha_{2}\beta_{2}}(\mu,\nu \mid \psi)$, attenuated by coefficients encoding the geometry of the relationship between the setting pairs $(\alpha_{1},\alpha_{2})$ and $(\beta_{1},\beta_{2})$. Since these coefficients take on both positive and negative values, some amplitudes cancel leaving the joint probability amplitude of the actual outcomes to match the observed probabilities of the EPRB correlations.

The way this has been constructed permits us to recognise this as an example of N\"{a}ger's internal cancelling paths, Fig.~\ref{fig:canpaths}, where $Y$ represents the intermediary state, $Z$ represents the final state and the cancelling path values are probability amplitudes between these states (each indexed by the measurement pair $\{(\alpha_{i},\beta_{j})\}$). And we can also recognise this as a type of Feynman path integral approach to the EPRB correlations, one that we can readily interpret along the lines of Wharton's retrocausal picture. Explicitly, we find the joint probability amplitude (\ref{eq:probamp}) to be a function of both the initial state $\psi$ and the final outcomes $a$ and $b$. We can imagine these amplitudes as being in some sense the beables that encode information about the future measurement settings $\alpha_{1}$ and $\beta_{1}$, along the lines of the treatment of the EPRB correlations in \citet{EvansPriceWharton}.

A further interesting feature of this model emerges when, instead of summing over the outcomes of the intermediary joint measurement, we simply model this measurement as a weak measurement (that is, $Y$ in Fig.~\ref{fig:canpaths} is a partially observed node).\footnote{There are similarities between this feature of the model and the weak measurement framework of the two-state vector formalism of \citet{AharonovVaidman90,AharonovVaidman91}.} When such a measurement is so weak that there is no measurement at all, we parallel the present case where we sum over the intermediary outcomes and the probability amplitudes cancel in such a way such that the final, actual outcomes match the observed probabilities of the EPRB correlations (and thus we have a standard violation of the CHSH inequalities). At the other extreme, when the intermediary measurement is a projective measurement, the intermediary outcomes are simply eigenstates of the corresponding measurement operator and we subsequently destroy the correlations in the final outcomes (and the CHSH inequalities will cease to be violated). Between these two extremes, a controllable partial intermediary observation would seem to partially prevent the precise cancelling of the probability amplitudes that gives us the EPRB correlations (and allow some sort of `tuning' of the violation of the CHSH inequalities).

Again, as I mentioned above, this is not as it stands a full blown retrocausal theory of quantum mechanics, it does however give a good idea of how an account of cancelling probability amplitudes constrained at two temporal boundaries can explain the observed correlations between measurement outcomes in an EPRB experiment.

\subsection{Mapping to a causal model}
\label{sec:retromodel}

It is interesting to consider how (or if) this retrocausal picture of the EPRB experiment could be cast as a causal model. One option for such a model would be the causal model represented in Fig.~\ref{fig:primretro} that we considered earlier. According to this model, the initial constraint, the preparation $P$, and the final constraints, the measurement settings $\bm{\alpha}$ and $\bm{\beta}$, act symmetrically on the inaccessible variable $\lambda$ to determine the beables that encode information about the outcomes $A$ and $B$. This is a DAG, as all edges are directed, and $\lambda$, by being spatiotemporally extended from the preparation to the measurement on each side of the experiment, embodies the nonlocality of the EPRB correlations.

However, when we begin to consider the exact nature of the measurement settings and outcomes in an EPRB experiment this raises a curious question about final boundary conditions. In a typical CHSH experiment, the measurement angle is chosen by manipulating a half wave plate which is then followed by a polarising prism and a pair of single photon detectors. The measurement setting is the choice of direction of the half wave plate (the effect of which is simply a unitary rotation of the polarisation state of the photon and is thus a completely reversible operation). The measurement outcome occurs when one of the detectors registers a photon, which indicates that a photon was present on the channel from the prism to that detector (and the probability distribution over the two outcome channels is conditional on the measurement setting). Thus there are aspects to the measurement outcome that are externally imposed---the constraint imposed by setting the half wave plate and the constraint imposed at the detector (that an integer number of photons occupy the outcome channel)---and there are aspects that are a result of the interrelation between the outcome and $\lambda$. The externally imposed constraints constitute the final boundary conditions and, along with the initial boundary conditions, partially determine the measurement outcome and, through this outcome, $\lambda$.

\begin{figure}
  \centering
  \begin{tikzpicture}[scale=1]
    \node (X) at (0,1) [circle,draw=black] {$\bm{\alpha}$};
    \node (A) at (1,2.5) [circle,draw=black] {A};
    \node (B) at (3,2.5) [circle,draw=black] {B};
    \node (Y) at (4,1) [circle,draw=black] {$\bm{\beta}$};
    \draw [-stealth] (X) -- (A);
    \draw [-stealth] (Y) -- (B);
    \node (L) at (2,0) [circle,draw=black] {$\lambda$};
    \node (P) at (2,-1.5) [circle,draw=black] {P};
    \draw [-stealth] (P) -- (L);
    \draw [-] (L) -- (A);
    \draw [-] (L) -- (B);
  \end{tikzpicture}
  \caption{A different set of causal (and other dependency) relations for a retrocausal model.}
  \label{fig:robretro}
\end{figure}
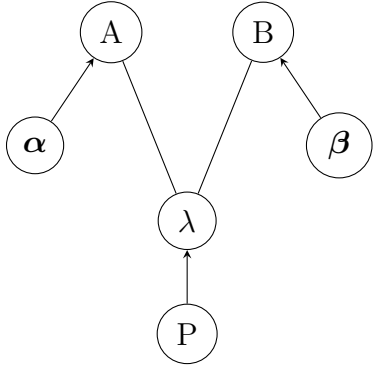

This suggests another way that we could cast the retrocausal picture as a causal model. Consider Fig.~\ref{fig:robretro}: the preparation variable $P$ is a direct cause of the hidden variable $\lambda$ and the measurement settings $\bm{\alpha}$ and $\bm{\beta}$ are direct causes of the outcome variables $A$ and $B$ and, in combination with the constraint imposed at each detector, provide a final boundary for $\lambda$. Because of this interrelation between the variables, it is a little unclear how the relationship should be represented; the quantum system between preparation and measurement is just the `solution' of a boundary value problem. Insofar as the boundaries play a causal role in determining the solution, $\bm{\alpha}$, $\bm{\beta}$ and $P$ are all partial causes of the solution, while $A$ and $B$ are a final manifestation of that solution (and $\lambda$ is some hidden element connecting the other variables); hence the undirected edges in Fig.~\ref{fig:robretro}. This model is, however, no longer a DAG\footnote{Fig.~\ref{fig:robretro} is obviously not directed, and arguably not acyclic as a result of the undirected edges.} and it remains to provide some account of how this model fits within the analysis of Wood and Spekkens, to which we now turn.

As we saw above, there are (at least) two types of responses that can be made as a consequence of the result that the three assumptions at the end of Sec.~\ref{sec:nofinetune} cannot be held simultaneously. The first response is the conclusion that Wood and Spekkens themselves reach, that the EPRB correlations resist any causal explanation and thus we should perhaps focus on amending the classical probability calculus that underpins Bayesian networks in order to accommodate quantum mechanics. The second response is to weaken the no fine-tuning assumption, and we have seen some different ways that we might do that according to \citet{Cartwright01}, \citet{EggEsfeld} and \cite{Naeger}.

%It would seem, at least superficially, that the causal model represented in Fig.~\ref{fig:robretro} might correspond more closely with the former response---at least insofar as it provides a model that goes beyond the standard causal model---and the causal model represented in Fig.~\ref{fig:primretro} might correspond more closely with the latter response---since it is a standard causal model that requires fine-tuning. It is unclear with which sort of model the current retrocausal picture matches, and it is certainly possible that in some sense the two blur together.

Let us consider the response aimed at weakening the no fine-tuning assumption. The line of argument that has helped us reach this point seems to endorse that we simply settle for the idea that the retrocausal picture of the EPRB experiment is an explicit example of fine-tuning whereby the tuning is a result of the sum-over-paths nature of the Feynman path integral approach. The subsequent unfaithfulness would be stable to disturbances and thus meets the criterion set out by N\"{a}ger as a counterpoint to Wood and Spekkens. Alternatively, rather than argue for stability of unfaithfulness, one could argue that the nature of the sum-over-paths explanation of the fine-tuning provides independent motivation for the violation of faithfulness, which would place this retrocausal picture in the same camp as the use of the quantum equilibrium hypothesis for Bohmian mechanics. According to such a view, the Feynman path integral approach would provide fundamental insight into the nature of reality and this is the foundation of the absence of signalling in the EPRB experiment.

But we could just as well interpret the retrocausal picture espoused here as endorsing that a causal explanation of the EPRB experiment, at least in the form of a DAG interpreted as a Bayesian network, is not possible, but perhaps for different reasons than Wood and Spekkens propose. To see this, consider the claim from \citet{HausmanWoodward}, based on an interventionist account of causation, that the nature of the EPRB correlations shows that $A$ and $B$ are not two distinct systems. As the two undirected edges of Fig.~\ref{fig:robretro} suggest, there are connections between parts of the system that do not allow for a `surgical' intervention to be made on one variable in isolation of any influence of the intervention on the other variables in the model---so-called `fat-handed' interventions. When this is the case DAGs are simply inapplicable because the causal Markov condition fails (in short, there cannot be a screening-off common cause) and \citet[p.~566]{HausmanWoodward} argue that the EPRB correlations are just such a case. Our retrocausal picture, particularly modelled as in Fig.~\ref{fig:robretro}, is consistent with this view.

There is an interesting take on this perspective, however. \citet{Hausman} argues that the lesson from the EPRB correlations is that it is possible for there to exist a (symmetric) mutual dependence between variables of a system that is not a causal dependence. Such variables, being mutually dependent, will have the same causes and effects. In the case of the EPRB correlations, this mutual dependence between variables is clearly a nomological dependence. According to \citet[p.~90]{Hausman}, ``Causal explanation is out of place because the separate measurements do not have independent causes. The modularity [absence of fat-handedness] that makes causal explanation appropriate does not obtain.''

\citet{Woodward14} makes a similar point. While arguing in favour of his interventionist framework in the context of so-called `causal exclusion' arguments, he suggests that it is possible that there be non-causal dependency constraints characterising the relationship between variables modelling some system. Such constraints are manifestations of a violation of what \citet[p.~14]{Woodward14} calls ``Independent Fixability'', whereby ``it is impossible\ldots for some combinations of values of variables to occur and \emph{a fortiori} for the variables to be set to those values by interventions''. This is precisely the case with the EPRB correlations, where some constraint is prohibiting, say, correlated or anti-correlated values of the outcome variables given certain measurement settings; this non-causal dependency constraint is just the mutual dependence or nomological constraint of Hausman. Woodward ultimately argues that we can still probe the causal relations of a system by interventions, but where we find non-causal dependency constraints such interventions must be understood as operating to set constrained variables only to combinations of values that obey the constraints. In other words, the constrained set of variables is intervened upon as a whole (or, in Hausman's terms, the mutually dependent variables have the same causes and effects).

In this context, we could then imagine the undirected edges of Fig.~\ref{fig:robretro} to be non-causal dependency (nomological) constraints that prohibit the independent fixability of the constrained variables. When we consider these undirected edges in this way, the distinction between the constrained variables $A$, $B$ and $\lambda$ collapses; these variables are intervened upon as a whole and have causes and effects as a whole. It is not immediately clear whether this sort of schema can be accommodated as a straightforward modification of standard causal modelling, or whether it can be accommodated at all. But I would like to proffer one suggestion concerning how we might think of the undirected edges of Fig.~\ref{fig:robretro}.

One way of thinking about causal relations from within a retrocausal picture is by considering causation to be `perspectival'.\footnote{See \citet{Price07}, with further discussion of the view in, for instance, \citet{Suarez07}, \citet{Silberstein08}, \citet{Ismael} and \citet{Evans15}.} The key to causal perspectivalism is to realise that we have a spatiotemporally constrained perspective within the `block universe' and, from this perspective, we have limited epistemic accessibility to other spatiotemporal regions; our notion of causation is a reflection of this. When we deliberate and act as causal agents, we do so to bring about events that occur in those inaccessible and unknowable regions of spacetime. In the current context, so long as at least some of those inaccessible regions of spacetime are in the past---as is the case for the hidden variables of the above models \citep[p.~13]{Evans15}---then we have retrocausal influences.

Thus we might imagine causal models to provide dependency relations between variables that we overlay with causal information that is a function of some set of temporally and epistemically directed interventions that we make on a system, and this is ultimately grounded in our perspective. The directed edges of a causal model, according to this view, would then represent the direction in which information is gained about a system---which is the standard Bayesian conception of causal models. We usually update information from past to future, so the directed edges are usually temporally directed. With regards to the undirected edges under consideration, it is not only the case that there is the potential for these edges to be directed in the reverse temporal direction, but depending upon the variables to which an agent can attribute values (i.e. the agent's perspective), the undirected edges have the potential to be directed in either temporal direction. Let me make this point more concretely.

If we were to obtain information concerning the measurement outcome $A$, but lacked information concerning the distant outcome $B$, we could tell a narrative regarding the influence of $A$ on $\lambda$, and the subsequent influence of $\lambda$ on $B$. We would represent this particular circumstance with corresponding causal arrows. Likewise, if we were to obtain equivalent information concerning the measurement outcome $B$, but lacked information concerning the distance outcome $A$, we could tell an alternative narrative regarding the causal influences in the system, from $B$ to $\lambda$ and from $\lambda$ to $A$. Neither would represent any objective causal structure of the system, only the causal structure we see from our perspective (whatever that happens to be). The objective structure would contain some non-causal, nomological dependence as depicted by the undirected edges of Fig.~\ref{fig:robretro}. This provides a novel conception of what causal models represent.

As one final suggestion, yet another perspective we could take on this retrocausal picture is that the sum over complex probability amplitudes introduces a nonclassical probability calculus over the classical probability structure of causal Bayesian networks (this is, of course, just standard quantum mechanics). This would align such a retrocausal account of quantum mechanics, at least in spirit, with the program advocated in \citet{LeiferSpekkens2013}. However, it is unclear how exactly one could represent the multitude of probability amplitudes that contribute to each path's probability neatly in a causal model.

\section{Conclusion}

I set out to enumerate the responses available to an advocate of retrocausal explanations of the EPRB correlations. Not all of these responses will be appealing, but the fate assigned to retrocausal explanations by Wood and Spekkens---that such explanations violate ``a principle that is at the core of all the best causal discovery algorithms'' and thus should be ruled out---seems much less strong in the face of these responses. One line of response is to explicitly disavow the sentiment that the faithfulness assumption is as sacred as suggested in order to maintain the causal modelling framework. I focussed, though, on a second line of response, in which faithfulness is violated in a principled or law based manner that is stable to background conditions, and which I take to have greater merit. However, whereas N\"{a}ger argues that his version of a stable violation of faithfulness by internal cancelling paths can save the central ideas of causal explanations, I consider a retrocausal version of internal cancelling paths.

The retrocausal picture I present, and the model of cancelling probability amplitudes, do not amount to a theory of quantum mechanics. This picture does, however, suggest an interesting direction for retrocausal models. In particular, understanding the connection between the weak measurement interpretation of the intermediary measurement in the model of Section~\ref{sec:cancelpath} and the two-state vector formalism of \citet{AharonovVaidman90,AharonovVaidman91} could provide a promising avenue of research.

I make the suggestion that, perhaps, the causal modelling framework is not sufficiently rich to accommodate the nature of retrocausal influences; certainly not of the sort I describe here. I think there is more to be done in this regard---for instance, is there an appropriate graphical representation that can capture the nuanced causal relationships of this picture? I hope to have provided here at least the beginnings of an exploration of this. Regardless of the success or otherwise of such a program, the door is certainly not closed on retrocausal approaches to quantum mechanics based on the framework of causal modelling.

\providecommand{\noopsort}[1]{}

\end{document}